\documentclass[11pt,a4paper]{article}

\usepackage{geometry}
\usepackage{amssymb,amsfonts,amsmath,stmaryrd}
\usepackage[export]{adjustbox}
\usepackage{enumerate,float,indentfirst}
\usepackage{color}
\usepackage{tikz}
\usepackage{tikz-cd}
\usetikzlibrary{arrows,snakes,backgrounds}
\usepackage{url}
\definecolor{darkblue}{rgb}{0.1,0.1,.7}
\usepackage[colorlinks, linkcolor=darkblue, citecolor=darkblue, urlcolor=darkblue, linktocpage]{hyperref}
\usepackage{textcomp}
\usepackage{subfig}
\usepackage{amsthm}
\usepackage{bbold}
\usepackage{mathtools}
\usepackage{multirow}
\usepackage{dsfont}
\usepackage{braket}
\usepackage{leftindex}
\usepackage[normalem]{ulem}
\usepackage[square, comma, compress,numbers]{natbib}
\usepackage{stackrel}
\usepackage[vcentermath]{youngtab}
\usepackage{caption}

\numberwithin{equation}{section}
\usepackage{amstext} 
\usepackage{array}   
\newcolumntype{L}{>{$}l<{$}} 
\newcolumntype{C}{>{$}c<{$}}

\def\p{\partial}

\newcommand{\td}[1]{\mathcal{#1}}
\newcommand{\veps}{\varepsilon}
\newcommand{\eps}{\epsilon}

\newcommand{\calV}{\mathcal{V}}
\newcommand{\calO}{\mathcal{O}}
\newcommand{\calL}{\mathcal{L}}

\newcommand{\ord}{\mathcal{O}}

\newcommand{\bz}{\bar{z}}
\newcommand{\bq}{\bar{q}}

\begin{document}

\vspace*{-.6in} \thispagestyle{empty}
\begin{flushright}
\end{flushright}
\vspace{1cm} {
\begin{center}
{\bf \Large On the Wilson–Fisher fixed point in the limit\\[3pt] of integer spacetime dimensions}
\end{center}}
\vspace{1cm}
\begin{center}
{Bernardo Zan}\\[2cm] 

Dipartimento di Fisica, Università di Genova and INFN, Sezione di Genova,\\ Via Dodecaneso 33, 16146, Genoa, Italy
\vspace{1cm}

\vspace{1cm}\end{center}

\vspace{4mm}

\begin{abstract}
The Wilson–Fisher fixed point defines a continuous family of interacting conformal field theories in non-integer dimensions. In integer dimensions, it is widely believed to lie in the same universality class as the critical Ising model. In this work, we revisit the identification between the Wilson–Fisher fixed point at integer dimensions and the Ising CFT. We argue that a literal equality between the two theories is incompatible with the emergence of Virasoro symmetry in two dimensions. Instead, we propose that the Ising model emerges only as a subsector of the Wilson–Fisher fixed point. We support this scenario through a detailed study of the two-dimensional $O(n)$ model and by examining operators transforming in irreducible representations of the orthogonal group whose multiplicities become negative for integer values of the spacetime dimension. Finally, we comment on the implications of these results for attempts to construct a $d=2+\eps$ expansion starting from exact two-dimensional data.
\end{abstract}
\vspace{.2in}
\vspace{.3in}
\hspace{0.7cm} Jan 2026

\newpage

\tableofcontents

\setlength{\parskip}{8pt}

\section{Introduction}

The textbook example of an interacting conformal field theory is the Wilson-Fisher (WF) fixed point \cite{Wilson:1971dc}. This fixed point can be obtained as the infrared fixed point of a scalar theory with quartic interactions in $d=4-\eps$ dimensions,
\begin{equation} \label{eq:WF_action}
	S  = \int d^{\,d} x\, \left[  \frac{1}{2}(\partial_\mu \phi)^2 + g \mu^{4-d} \phi^4 \right]\,.
\end{equation}
A study of the beta function of this theory shows that it admits a zero at $g \sim \ord(\eps)$, meaning that the theory is weakly coupled for $0<\eps \ll 1$. This fixed point defines the WF conformal field theory. Historically, the importance of the WF fixed point lies in the fact that it provided the first example of an interacting fixed point that is accessible within perturbation theory.

As $\eps$ becomes $\ord(1)$, the WF fixed point becomes strongly coupled, and analytic control over the theory is lost. Nevertheless, the fixed point is still believed to exist. The theory possesses an internal $\mathbb{Z}_2$ symmetry, and one therefore expects it to lie in the same universality class as other critical theories with $\mathbb{Z}_2$ symmetry in the same number of spacetime dimensions. In particular, in integer dimensions $d=2$ and $d=3$, the WF fixed point is expected to describe the critical Ising model. This expectation is supported by computations of the scaling dimensions of the lightest operators. Working in an expansion in $\eps$ to very high order,\footnote{See e.g. \cite{Schnetz:2022nsc} for the seven-loop results.} resumming the resulting asymptotic series and then setting $\epsilon \to 1$ \cite{Kompaniets:2017yct,Bonanno:2022ztf} or $\epsilon \to 2$ \cite{Kompaniets:2017yct}, one finds very good agreement with rigorous non-perturbative numerical results in $d=3$ \cite{El-Showk:2013nia,Kos:2016ysd,Chang:2024whx} and with exact results in $d=2$ \cite{onsager,Yang:1952xj,Belavin:1984vu}.

The main objective of this work is to clarify the precise connection between the WF fixed point in integer dimensions and the Ising CFT, with particular emphasis on the case $d=2$. We will show that the naive expectation that the $\eps \to 2$ limit of the WF fixed point coincides with the Ising model is incompatible with the emergence of infinitely many conserved currents in two dimensions, which follows from Virasoro symmetry. Motivated by a two-dimensional model whose spectrum is exactly known, we propose a more general alternative scenario that avoids this paradox. According to this scenario, the two (three) dimensional Ising CFT emerges only as a subsector of the WF fixed point in the $d \to 2$ ($d \to 3$) limit.

The WF CFT is naturally formulated as an expansion around $d=4$. However, given its connection with the Ising model in integer dimensions, a natural question is whether the WF fixed point can instead be described in an expansion around $d=2$, using the exact conformal data of the two-dimensional Ising model, or around $d=3$, using the high-precision numerical results available for the three-dimensional Ising model. We will discuss the implications of our scenario for this question. The central point is that the Ising CFT constitutes only a subsector of WF in integer dimensions. As a result, it appears unlikely that the conformal data of the full theory in $d=2+\eps$ can be obtained starting solely from a subsector of the theory in $d=2$. More concretely, we show that when one starts from the crossing equations of a set of operators in $d=2$ and deforms to $d=2+\eps$, new operators can appear with coefficients of order $\ord(1)$.

The paper is organized as follows. In Section \ref{sub:the_problem}, we define the issue that arises when attempting to identify the $d \to 2$ limit of the WF fixed point with the $d=2$ Ising CFT. In Section \ref{sec:2d_On}, we then turn to the two-dimensional $O(n)$ model for $n \in \mathbb{R}$ as a toy model for the WF fixed point in non-integer spacetime dimensions. In Section \ref{sec:WF_2d}, we look for evidence supporting this scenario directly within the WF fixed point by discussing operators transforming in representations of the orthogonal group that have negative multiplicity at $d=2$. Additional details about the two-dimensional $O(n)$ model are collected in Appendix \ref{app:on}.

\subsection{The problem} \label{sub:the_problem}

We consider the WF fixed point in $d$-dimensional Euclidean space and treat the number of spacetime dimensions as a non-integer parameter, $d \in \mathbb{R}$; only at the end will we take the limit in which $d$ approaches an integer. We make two assumptions about the theory, which are valid in the perturbative regime $0<\eps \ll 1$ and which we assume continue to hold more generally for $\eps \sim \ord(1)$:
\begin{enumerate}
\item the theory exists and is well defined for $d \in \mathbb{R}$, $2\le d\le 4$;
\item the conformal data of the theory depends analytically on $d$ for $2<d<4$.
\end{enumerate}

Throughout this work, we restrict our attention to the range $2 \le d \le 4$.\footnote{The upper critical dimension of the Ising model is $d=4$, while its lower critical dimension is $d=1$ \cite{Goldenfeld:1992qy}. This means that a continuation of the Ising model down to $d=1+\eps$ is still an interesting theory \cite{Bruce:1981lse}. We will not discuss the range $1<d<2$ in this work, but we mention that a bootstrap study of this range of dimensions gave inconclusive results \cite{Golden:2014oqa}.} By well-defined we mean that the usual axioms of Euclidean CFTs are satisfied, such as conformal invariance, locality, the existence of an OPE and crossing symmetry. We do not, however, assume that the theory is unitary,\footnote{Throughout this work, we use the terms unitarity and reflection positivity interchangeably as we work in Euclidean signature.} as it was shown in \cite{Hogervorst:2014rta,Hogervorst:2015akt} that a theory in non-integer dimension is nonunitary. This lack of unitarity is responsible for many of the subtleties that we encounter.
The second assumption implies that we do not expect the theory to jump suddenly as we vary the value of $d$. More precisely, we assume that operator dimensions, their multiplicities and their OPE coefficients are analytic functions of $d$. %

The statement that the WF fixed point describes the Ising model for $d=2,3$ can be interpreted in different ways. We now assume the most literal version of this statement, namely that the conformal data of the WF CFT at $d=2,3$ coincides exactly with that of the Ising model in the same number of dimensions. We will show that this interpretation cannot be correct by examining the case $d=2$.

Since the theory is a local CFT, it possesses a conserved and traceless energy-momentum tensor $T^{\mu \nu}$ for any value of $d$,
\begin{equation}
\partial_\mu T^{\mu \nu}=0,.
\end{equation}
Generically, there are no additional conserved currents, as can be verified in $d=4-\epsilon$ dimensions.

In contrast, the two-dimensional Ising model has infinitely many conserved currents due to Virasoro symmetry. For example, a spin-4 conserved current in two dimensions is
\begin{equation}
	\td{T}_4 \sim \left(L_{-4}-\frac{5}{3}L_{-2}^2 \right)\ket{0}
\end{equation}
where, from now on, we denote two-dimensional operators with a curly font. Although this operator is a Virasoro descendant, it is a primary with respect to the global conformal group, since $L_1 \td{T}_4 =\overline{L}_1 \td{T}_4 =0$. The operator $\td{T}_4$ is conserved, $\bar{\partial} \td{T}_4(z)=0$, as can be seen from
\begin{equation}
 	\overline{L}_{-1} \td{T}_4 \ket{0}= 0\,,
\end{equation} 
and similar statements hold for the antiholomorphic operator $\overline{\td{T}}_4$. 
Similarly, there exist infinitely many higher-spin conserved currents that are Virasoro descendants but global primaries, for instance a spin-$6$ one, two spin-$8$ ones and so on. This structure can be seen by decomposing the vacuum character of the Ising model into characters of global primaries.

In the WF CFT for $d>2$, the absence of higher-spin conserved currents implies that the lightest $\mathbb{Z}_2$ even spin-4 operator $T_4^{\mu \nu \rho \sigma}$ is not conserved for generic $d$. However, if we assume that this operator becomes $\td{T}_4$ (and $\overline{\td{T}}_4$) as $d \to 2$, then it must become conserved at $d=2$. We can therefore write for $d-2\ll 1$
\begin{equation}
\partial_\mu T_4^{\mu \nu \rho \sigma} = (d-2)^\alpha W^{\nu \rho \sigma}+\ldots\,.
\end{equation}
The operator $W$ is a spin-3 descendant of $T_4$, whose scaling dimension is fixed in terms of the dimension of $T_4$. The exponent $\alpha$ needs to be positive.

Now comes the problem with interpreting the $d\to2$ limit of WF as being exactly the 2d Ising CFT: if we assume that $T_4 \to \td{T}_4$ when $d\to2$,\footnote{In the $d\to 2$ limit, $T_4$ reduces to the sum of the holomorphic and antiholomorphic components $\td{T}_4$ and $\overline{\td{T}}_4$. For notational simplicity we just write the holomorphic component.} then this fixes $W \to \td{W}$, where $\td{W}$ is a $\Delta = 5$, spin 3 operator. If the conformal data behaves analytically in $d$, this operator cannot simply disappear, and we must end up with an operator with $\Delta = 5$ and $\ell = 3$ in the $d\to 2$ limit. We can argue the presence of this operator also by going from $d=2$ to $d=2+\eps$. The global primary $\td{T}_4$ is part of a short multiplet at $d=2$, while $T_4$ is part of a long multiplet for $d>2$. The mechanism by which such phenomenon can occur is multiplet recombination \cite{Rychkov:2015naa}, i.e. by recombining with another operator which is a global primary at $d=2$ and becomes a descendant at $d>2$. This operator must have the right quantum numbers at $d=2$, which are $\Delta =5$ and $\ell = 3$.

Here is the paradox: the conformal data of WF in the limit $d\to2$ cannot match that of the 2d Ising model, because the spectrum of the 2d Ising CFT contains no operator with $\Delta = 5$ and $\ell = 3$!\footnote{The author has first learned about this issue from Leonardo Rastelli. This has also been subsequently discussed in \cite{Zhou:2022pah}.}
It is straightforward to check the absence of such operator by expanding the torus partition function of the theory. Indeed, the two-dimensional Ising contains three Virasoro primaries, $\mathds{1}$, $\sigma$ and $\veps$. A descendant with dimension $\Delta = 5$ and spin $\ell=3$ has holomorphic and antiholomorphic weights $(h,\bar{h}) = (4,1)$ (or vice versa), and can therefore only be a Virasoro descendant of the identity. However, the level one descendant of the identity is null, $L_{1}\mathds{1}=\bar{L}_{1}\mathds{1} = 0$, so no such operator appears in the theory.

We have shown this for the lowest higher-spin current, but the same argument applies to the higher ones. For example, the lightest $\mathbb{Z}_2$ even spin-6 operator should become conserved at $d=2$. However, the Ising model has no operator with $\Delta = 7$ and $\ell = 5$ that could play the role of its descendant for $d \gtrsim 2$. The conclusion is the same: identifying the full WF spectrum in the $d \to 2$ limit with that of the Ising model is not consistent.

What are the possible resolutions of this paradox? One option is to abandon the idea that, for integer dimensions, the WF CFT describes the Ising model. However, this would leave us with the task of explaining why the $\epsilon$ expansion gives accurate results for the critical exponents of the Ising model in both $d=2$ and $d=3$ \cite{Kompaniets:2017yct}. Another possibility is that, as $d$ approaches an integer, we recover the Ising model together with an additional decoupled sector. This is similar to what happens in nonlocal theories with power-law interactions, when these interactions decay sufficiently fast \cite{Behan:2017dwr,Behan:2017emf,Behan:2018hfx,Behan:2025ydd}.

We will argue that a more generic scenario is that the 2d Ising model appears only as a subsector of WF in the $d\to2$ limit. In the example above, $\td{T}_4$ is part of the Ising subsector, while $\td{W}$ is not. This scenario is motivated by the study of certain two-dimensional CFTs in which the same phenomenon occurs \cite{Gorbenko:2020xya,Gabai:2024puk}. Indeed, it has already been pointed out that when a theory has a symmetry group with an integer parameter $x$ that can be continued to real values, there is a difference between considering the theory strictly at $x=x^* \in \mathbb{N}$ and taking the $x\to x^*$ limit \cite{Cardy:2013rqg,Hogervorst:2015akt}.\footnote{See also \cite[Section 2.5]{Henriksson:2025kws} for a similar discussion and a collection of references.} Here we put forward a specific mechanism through which this phenomenon can manifest itself. In this work we mainly focus on the limit $d\to 2$: here the appearance of infinitely many conserved quantities makes the paradox manifest. However, the issue and the scenario we propose to resolve it also apply to $d \to 3$.

\section{The two-dimensional $O(n)$ CFT as a toy model} \label{sec:2d_On}

In order to see in a hands-on way what happens in a similar situation in a strongly coupled theory, we now consider the two-dimensional $O(n)$ model. Microscopically, this can be defined as a statistical mechanics model with $n$-component spins $\vec{S}_i$ living on the sites of a lattice and interacting in an $O(n)$-symmetric way \cite{Stanley:1968ef}. This definition makes sense for $n$ being a positive integer. Through a high-temperature expansion, the same model can also be written as a loop model, which allows for an extension to real values of $n$ \cite{Domany:1981fg}. For $-2<n\le2$ the model is critical at a real value of the temperature \cite{Nienhuis:1982fx}, and the corresponding CFT spectrum is known exactly \cite{diFrancesco:1987ses}.

The parameter $n$ in the $O(n)$ model will play the role of $d$ in the WF CFT:
when defined as a spin model with $n \in \mathbb{Z}_{>0}$, the theory is unitary; in its loop formulation, however, unitarity is lost \cite{Binder:2019zqc}. This is the same pattern as that observed in theories in integer and non-integer dimensions.

We are interested in the continuum limit of the critical $O(n)$ model. The starting point for our analysis is the torus partition function of the $O(n)$ model \cite{diFrancesco:1987ses,Read:2001pz}
\begin{equation} \label{eq:ZOn}
	Z_n = 
\frac{1}{\eta(q)\eta(\bar{q})} \left\{
\sum_P q^{x_{e_0 + 2P, 0}} \bar{q}^{\bar{x}_{e_0 + 2P, 0}}
+ \sum_{\substack{M > 0, N > 0, P:\\ N|M,\, P \wedge N = 1}} 
\Lambda_{M,N}\, q^{x_{2P/N,\, M/2}} 
\bar{q}^{\bar{x}_{2P/N,\, M/2}}
\right\}\,.
\end{equation}
Here $N|M$ means that $N$ divides $M$, and $P \wedge N$ denotes the greatest common divisor of $P$ and $N$, with the convention $0 \wedge N = N$. $\eta$ is the Dedekind eta function. The functions $x$ and $\bar{x}$ are defined by
\begin{equation}
\begin{aligned}
	x_{e,m} &= \frac{1}{4} \left( \frac{e}{\sqrt{g}} + m\sqrt{g} \right)^{2}, \\
\bar{x}_{e,m}&= \frac{1}{4} \left( \frac{e}{\sqrt{g}} - m\sqrt{g} \right)^{2}\,
\end{aligned}
\end{equation}
with the `coupling' $g$ related to $n$ by
\begin{equation}
	n = -2 \cos \pi g\,, \qquad g \in[1,2]\,,
\end{equation}
and the `background charge' $e_0$ by
\begin{equation}
e_0 = g-1\,.
\end{equation}
Finally, an explicit form of $\Lambda_{M,N}$ can be found in \eqref{eq:Lambda}.

Knowledge of the torus partition function determines the spectrum of the theory for generic values of $n$. In particular, we can read off the central charge,
\begin{equation}
	c = 1-\frac{6 e_0^2}{g}\,.
\end{equation}
For generic $n$ the theory is non unitary. This can be seen very explicitly by computing correlation functions and observing that the theory is a logarithmic CFT \cite{Gorbenko:2020xya,Nivesvivat:2020gdj}, which implies the presence of negative-norm states and hence non-unitarity.

For special values of $n$, the $O(n)$ partition function \eqref{eq:ZOn} reduces to well-known unitary CFTs. For $n=1$ it coincides with the partition function of the Ising model,
\begin{equation}
	Z_1 = \chi_0(q)\chi_0(\bq)+\chi_{1/16}(q)\chi_{1/16}(\bq)+\chi_{1/2}(q)\chi_{1/2}(\bq)\,,
\end{equation}
where explicit formulas for the Virasoro characters $\chi_h(q)$ at $c=1/2$ can be found for example in \cite[Chapter 8]{DiFrancesco:1997nk}. For $n=2$ the partition function describes the $O(2)$ model, which is the $c=1$ theory at the BKT radius \cite{Dijkgraaf:1987vp},
\begin{equation}
	Z_{2} = \frac{1}{\eta(q)\eta(\bar{q})} \sum_{l,k=-\infty}^\infty q^{\left(l+\frac{k}{4} \right)^2} \bq^{\left(l-\frac{k}{4} \right)^2}\,.
\end{equation}

By now, it should be clear why the $O(n)$ CFT can serve as a toy model for the WF fixed point. Both models form a one-parameter family of CFTs, with the parameter controlling the symmetry of the theory. For generic values of this parameter the theories are non unitary, while for special values they are expected to reduce, in some sense, to unitary CFTs. In particular, both the $n\to 1$ limit of the $O(n)$ model and the $d\to 2$ limit of the WF fixed point should describe the two-dimensional Ising model in some appropriate sense. We will therefore use the $n\to 1$ limit of the $O(n)$ model as a toy model for the $d\to 2$ limit of the WF fixed point. The clear advantage is that the spectrum of the $O(n)$ model in two dimensions is known exactly, so we can take the $n \to 1$ limit in a controlled way.

In order to claim, for example, that the $n \to 1$ limit of the $O(n)$ model is the Ising model, it is not sufficient to compare partition functions. We must also study correlation functions, which we now do.

\subsection{Correlation functions}

We now study some correlation functions of the $O(n)$ model. While the spectrum has been known since \cite{diFrancesco:1987ses}, the computation of OPE coefficients proves trickier; a method to obtain OPE coefficients of generic operators was developed in \cite{Grans-Samuelsson:2021uor,Nivesvivat:2023kfp}. Here we focus only on a few four-point functions that involve degenerate operators, i.e. operators in the Kac table. These four point functions satisfy BPZ differential equations \cite{Belavin:1984vu}, which considerably simplify the analysis.

It is convenient to parametrize $n$ as
\begin{equation}
	n = 2 \cos \left(\frac{\pi }{m}\right)\,.
\end{equation}
In this parametrization the central charge is
\begin{equation}
	c = 1-\frac{6}{m(m+1)}\,,
\end{equation}
which coincides with that of the (generalized) minimal models $\mathcal{M}_{m,m+1}$ \cite{Belavin:2005yj,Zamolodchikov:2005fy}. The Ising limit $n \to 1$ corresponds to $m \to 3$, while the limit $n \to 2$ corresponds to $m \to \infty$. We parametrize scaling dimensions as
\begin{equation}
	h_{r,s} = \frac{((m+1)r-m s)^2-1}{4m(m+1)}\,.
\end{equation}
An operator with dimension $h_{r,s}$, where $r$ and $s$ are positive integers, is a degenerate operator. From the Virasoro algebra it follows that such an operator has a descendant of zero norm at level $rs$.

At this point we need to stress a subtle feature of non-unitary theories. We call zero norm operators those whose two-point function with themselves vanishes, and null operators those whose every correlation function vanishes, so that they can be consistently modded out of the theory. While null operators always have zero norm, the converse is not necessarily true. In a unitary theory, where we have a well-defined non-negative norm, zero norm implies the vanishing of an operator, so it is null and decouples from the theory. Here, however, we are dealing with a non-unitary theory, and this implication does not necessarily hold. Therefore not every zero-norm operator is necessarily null. Following \cite{Gorbenko:2020xya}, we adopt the following criterion. If a zero norm descendant is absent from the partition function for generic values of $n$, we regard it as not part of the theory and therefore null.\footnote{Compare this to the case where the Kac labels of an operator become integers only for a specific value of $n$. In that situation the descendant operator appears in the partition function at that special value of $n$. It has zero norm but is not null.}

Let us consider the energy operator $\veps$, the lightest $O(n)$ singlet above the identity. This operator is a scalar and has weights $(h_{1,3},h_{1,3})$. This implies that one of its descendants at level three has zero norm. Inspecting the partition function, we see that only two operators with dimensions $(h_{1,3}+3,h_{1,3})$ are present, out of the three possible descendants. According to our criterion, this means that the energy operator $\veps$ has a null descendant at level three.

Let us now consider the four-point function of $\veps$. The fact that the level three descendant is null implies that the fusion rule of $\veps$ is
\begin{equation}
	\veps \times \veps = \mathds{1}+\veps+\veps'\,,
\end{equation}
where we have introduced $\veps'$, the subleading scalar singlet under $O(n)$, with dimensions $(h_{1,5},h_{1,5})$.\footnote{The relation between having a null descendant and the fusion rule can be seen e.g. by considering the four point function $\braket{\veps(0)\veps(z,\bz)\calO_1(1)\calO_2(\infty)}$. The presence of a null descendant at level three means that the correlation function satisfies a homogeneous third degree differential equation, the BPZ differential equation \cite{Belavin:1984vu}. The three independent solutions are the Virasoro blocks for the exchanged operators, and their dimensions can be read from expanding these solutions to small $z,\bz$. See Appendix \ref{app:on} for more details.} Correlation functions of $\veps$ for generic $n$ can therefore be expressed as
\begin{equation}\label{eq:eeee_n}
	\braket{\veps(0)\veps(z,\bz)\veps(1)\veps(\infty)} = \calV_0(z) \bar{\calV}_0(\bz)+\lambda_{\veps \veps \veps}\calV_\veps(z) \bar{\calV}_\veps(\bz)+\lambda_{\veps \veps \veps'}\calV_{\veps'}(z) \bar{\calV}_{\veps'}(\bz)\,,
\end{equation}
where $\calV_\calO$ is the Virasoro block associated with the exchange of the operator $\calO$, and $\lambda_{ijk}$ are OPE coefficients. Using the BPZ differential equations, one can determine the Virasoro blocks, and crossing symmetry then fixes the values of the OPE coefficients; see Appendix \ref{app:on} for more details. These data necessarily coincide with those of generalized minimal models.

Once the Virasoro blocks and OPE coefficients are known for generic values of $n$, we can take the $n \to 1$, or $m \to 3$, limit of this correlation function, obtaining
\begin{equation}\label{eq:eeee_is}
	\lim_{n\to 1} \braket{\veps(0)\veps(z,\bz)\veps(1)\veps(\infty)}= \left| \frac{1}{z}+\frac{z}{1-z} \right|^2\,.
\end{equation}
This is precisely the four-point function of the energy operator in the Ising model.

\subsubsection{Correlation functions of non-Ising operators}
\label{ssub:correlation_functions_of_non_ising_operators}

So far everything looks consistent with the most straightforward interpretation of the $n\to 1$ limit of $O(n)$ as giving the Ising model. This is because we have considered the correlation function of the energy operator in $O(n)$, which becomes the energy operator of the Ising model when $n\to 1$. However, we will now study correlation functions of other operators that, for $n\to1$, we do not expect to appear in the Ising model. For example, the $O(n)$ model at generic $n$ has a conserved current that transforms in the two-index antisymmetric representation of $O(n)$. The Ising model clearly has no continuous global symmetry, since its only global symmetry is $\mathbb{Z}_2$. In the $n\to 1$ limit the current operator disappears from the torus partition function because its multiplicity vanishes. Nevertheless, one can still compute its correlation functions at generic $n$ and then take the limit $n \to 1$ to obtain a non-trivial observable. One can go further and consider operators that transform in irreducible representations of $O(n)$ whose dimension is negative at $n=1$, so that their multiplicity is negative in this limit. 

\paragraph*{An operator with vanishing multiplicity} 
Let us consider the current operator, which arises in \eqref{eq:ZOn} from the $P=0$ term in the first sum (after expanding the Dedekind eta function) and from the $M=2$, $N=2$, $P=\pm 1$ term in the second sum,
\begin{equation}
	Z_n = (q \bq)^{-c/24} \left[ \ldots+\frac{n(n-1)}{2} \left( q  + \bq  \right) + \ldots\right]\,.
\end{equation}
These operators can be identified with the currents $J$ and $\bar{J}$ because they have dimensions $(1,0)$ and $(0,1)$, and their multiplicity matches that of an operator in the two-index antisymmetric representation of $O(n)$, as expected. The correlation function $\braket{\veps \veps J J}$ can be computed for generic $n$, and one finds that it does not vanish in the limit $n \to 1$, see Appendix \ref{app:on} for more detailed expressions. 
This is the first of several examples where we will see that the $n \to 1$ limit of the theory is larger, in the sense that it has more non-trivial observables, than just the Ising model.

One might hope that the $n \to 1$ limit of the theory consists of two decoupled sectors, one being the Ising model and the other containing the remaining operators, including the current $J$. This is too optimistic. For instance, the $n \to 1$ limit of $\braket{\veps \veps J J}$ does not factorize into a product of two-point functions,
\begin{equation}
	\lim_{n \to 1} \braket{\veps \veps J J}  \neq \lim_{n \to 1} \braket{\veps \veps} \braket{J J}\,.
\end{equation}

\paragraph*{An operator with negative multiplicity}
The same phenomenon can be observed for other operators. Consider for example the term in the partition function
\begin{equation}
Z_n = (q \bq)^{-c/24} \left[ \ldots+\frac{n(n-2)(n+2)}{3} \left( q^{h_{\frac 32, -\frac 23}}  \bq ^{h_{\frac 32, \frac 23}} +q^{h_{\frac 32, \frac 23}}  \bq ^{h_{\frac 32, -\frac 23}}  \right) + \ldots\right]
\end{equation}
which comes from the $P=\pm1$, $M=N=3$ term in \eqref{eq:ZOn}. We denote these operators by $Y$. Their multiplicity is that of an operator in the representation of $O(n)$ associated with the Young tableau $(2,1)$. Although the multiplicity of this operator is negative at $n=1$, the four-point function $\braket{\veps \veps Y Y}$ can be computed in the $n \to 1$ limit, see again Appendix \ref{app:on}, and it is non zero. Since this operator does not make sense as an $O(n)$ representation strictly at $n=1$, this correlation function must belong to a theory that is larger than the Ising model. In this case as well we find that the four-point function does not factorize,
\begin{equation}
	\lim_{n\to 1} \braket{\veps \veps Y Y} \neq \lim_{n\to 1} \braket{\veps \veps}\braket{ Y Y}\,.
\end{equation}

\paragraph*{An operator with positive multiplicity}
A reader might complain that, by studying the $n \to 1$ limit of correlation functions of operators whose multiplicity either vanishes or becomes negative in this limit, we are simply looking for trouble. In order to avoid these problems, is it sufficient to restrict attention to operators in representations of $O(n)$ that behave well in the $n \to 1$ limit? Unfortunately, the answer is no. This can be seen either by direct inspection or from the fact that operators with negative multiplicity, such as $Y$, must cancel against operators with positive multiplicity in the $n\to1$ limit in order to reproduce the Ising partition function, which only contains operators with positive multiplicity.

An example of this is the operator arising from the $M=N=1$, $P=\pm1$ term in \eqref{eq:ZOn}. It transforms in the vector representation of $O(n)$, and therefore has multiplicity $n$. We denote it by $V$, and it has weights $(h_{1/2,-2},h_{1/2,2})$,
\begin{equation}
Z_n = (q \bq)^{-c/24} \left[ \ldots+n \left( q^{h_{ 1/2, -2}}  \bq ^{\,h_{ 1/2, 2}} +q^{h_{ 1/2, 2}}  \bq ^{\,h_{ 1/2, -2}} \right) + \ldots\right]\,.
\end{equation}
The fact that this operator does not appear in the Ising model follows from its weights, which in the $n\to 1$ limit become $(\tfrac{21}{16},\tfrac{5}{16})$; no operator with these weights exists in the Ising model. As in the previous cases, the corresponding four-point function is non-trivial and does not factorize in the $n\to 1$ limit.

Notice that, as $n\to1$, the dimensions of $V$ and $Y$ both approach $(\tfrac{21}{16},\tfrac{5}{16})$. While $Y$ has multiplicity $-1$, $V$ has multiplicity $+1$, so they cancel each other in the $O(n)$ partition function in the $n\to 1$ limit.\footnote{There is another descendant operator whose dimension becomes $(\tfrac{21}{16},\tfrac{5}{16})$ when $n\to1$, but its multiplicity vanishes in this limit.}$^,$\footnote{The four-point functions $\braket{\veps\veps V V}$ and $\braket{\veps\veps Y Y}$ actually coincide in the $n\to1$ limit, which suggests that the two operators form a logarithmic multiplet at $n=1$.}

\paragraph*{The general picture}

We have now seen in several examples that the $n\to1$ theory has more observables, in the sense of non-trivial correlation functions, than the Ising CFT: as an example, correlation functions including the vector operator $V$, which we have just seen, are well defined CFT correlation functions, but have nothing to do with the Ising CFT correlation functions.
The results point to a scenario in which, as we take the limit $n\to1$, the theory develops a subsector that reproduces the Ising model. Correlation functions of operators in this subsector exactly match Ising CFT correlators, so this subsector is unitary \cite{Gorbenko:2020xya}. At the same time, the full theory contains infinitely many additional operators that do not belong to this subsector and whose correlation functions are non trivial but do not admit an interpretation in terms of the Ising CFT. The requirement that the Ising subsector is closed means that the OPE of Ising operators can only contain operators in the Ising subsector.\footnote{A more complicated scenario might be like that found in \cite{Gabai:2024puk,Gabai:2024qum}, where a theory in some limit develops a (fermionic) Ising subsector in which operators belonging to the Ising sector generate operators outside of Ising in their OPE. However, many correlation functions of such operators vanish, so that the Ising sector still reproduces the Ising correlators. We do not consider this scenario as it requires an exotic global symmetry that is absent from the theories appearing in this work.}

Let us emphasize once more that the torus partition functions of the two theories agree exactly. This implies that the total multiplicity of non-Ising operators must vanish for every choice of Virasoro weights. In the case at hand, non-Ising operators either have zero multiplicity individually or operators with positive and negative multiplicities cancel each other.\footnote{On a similar note, the work \cite{Cao:2023psi} considered the continuation of $O(n)$ theories (in general dimensions) to non-integer $n$. In the limit of $n$ becoming an integer, it argued for $O(n)$ characters of poorly behaved representations to cancel with other representations. The two-dimensional $O(n)$ model discussed here is an explicit example where this phenomenon occurs. See also \cite{Antunes:2024mfb}, for a similar discussion for the permutation group.}

Operators that transform in $O(n)$ representations that are ill defined at $n = 1$ (because they have negative or vanishing multiplicity) cannot belong to the unitary subsector. However, there is no way to determine a priori whether an operator that transforms in a representation that survives as $n\to1$ lies in the Ising subsector or not. Examples include the spin field $\sigma$ and the operator $V$ discussed above. Both transform in the vector representation of $O(n)$, but only one of them belongs to the Ising subsector.

\subsection{From $n=1$ to $n=1.01$} \label{sec:O(n=1.01)}
Another question we will discuss later in Section \ref{sec:d=2.01} is whether knowledge of a theory at integer dimensions, e.g. the Ising model at $d=2$, is enough to obtain observables for the WF fixed point in $d=2+\eps$ dimensions.
This is another instance in which the $O(n)$ model can serve as a toy model for the WF fixed point. In analogy with this question, one can ask whether the spectrum of the two-dimensional $O(n)$ model at $n=1+\mathcal{O}(\delta)$ can be recovered from the two-dimensional Ising spectrum.

Let us again consider the four-point function of the energy operators for generic $n$, \eqref{eq:eeee_n}, which reduces to the Ising correlation function when $n\to1$. At first sight the $n \to 1$ limit appears to behave perfectly well. Recall that the fusion rule in the Ising model is
\begin{equation}
	\veps \times \veps = \mathds{1}\,,
\end{equation}
so we might be tempted to conclude that the exchanged operators $\veps$ and $\veps'$ have decoupled from the four point function $\braket{\veps\veps\veps\veps}$. We now show that the situation is actually not as straightforward.

Expanding the OPE coefficients \eqref{eq:eps_OPE} near $m=3+\delta$, we find
\begin{equation}
\begin{aligned}
	\lambda_{\veps\veps\veps}^2 &= \frac{\Gamma \left(\frac{3}{4}\right)^4}{\Gamma \left(\frac{1}{4}\right) \Gamma \left(\frac{5}{4}\right)^3}\,\delta ^2 +\mathcal{O}(\delta^3)\,,\\
	\lambda_{\veps\veps\veps'}^2 &= -\frac{\Gamma \left(-\frac{1}{4}\right)^3 \Gamma \left(\frac{15}{4}\right)}{3025\, \Gamma \left(-\frac{11}{4}\right) \Gamma \left(\frac{5}{4}\right)^3} +\mathcal{O}(\delta)\simeq -0.230 + \mathcal{O}(\delta)\,.
\end{aligned}
\end{equation}
While $\lambda_{\veps\veps\veps}$ vanishes as expected, $\lambda_{\veps\veps\veps'}$ does not. How, then, do we end up with only the Virasoro block of the identity in \eqref{eq:eeee_is}? Even more striking is the fact that $\lambda_{\veps\veps\veps'}$ becomes imaginary.

With the situation in non-integer dimensions in mind, it is more illuminating to consider the expansion of $\braket{\veps\veps\veps\veps}$ in global conformal blocks. We write
\begin{equation} \label{eq:global_exp}
\braket{\veps \veps \veps \veps}  = \frac{1}{(z \bz)^{\Delta_\veps}}\sum_{ \calO } C_{\Delta_{\calO},\ell_{\calO}} F_{\Delta_{\calO},\ell_{\calO}} (z,\bz)
\end{equation}
where $F_{\Delta,\ell}$ are the global conformal blocks in two dimensions \cite{Dolan:2000ut}
\begin{align}
 F_{\Delta,\ell}(z,\bz) &= \frac{1}{1+\delta_{\ell,0}}\left[ \kappa_{\Delta+\ell}(z)\,\kappa_{\Delta-\ell}(\bar z)
 + \kappa_{\Delta-\ell}(z)\,\kappa_{\Delta+\ell}(\bar z) \right]\,, \\
\kappa_\alpha(z)
&\equiv z^{\alpha/2}\,
{}_2F_1\!\left(\frac{\alpha}{2},\frac{\alpha}{2};\alpha;z\right) \,.
 \end{align} 
This expansion allows us to read off the OPE coefficients $C_{\Delta, \ell}$ of global primaries. A natural expectation would be that, since at $n=1$ we only have global primaries that are descendants of the identity in the $\veps \times \veps$ OPE, the coefficients $C_{\Delta,\ell}$ vanish in the $n\to1$ limit whenever $\Delta$ is not an integer, while they could remain non-zero when $\Delta$ is an integer.

This expectation turns out to be incorrect. In the $\delta\to 0$ limit we find many operators with $\Delta = \mathbb{Z}+\tfrac{1}{2}$ whose OPE coefficients are of order $\mathcal{O}(1)$. However, some of these coefficients are negative and cancel exactly against others in the limit $\delta \to 0$, so that the limit reproduces the correct answer for $\braket{\veps\veps\veps\veps}$. This is completely analogous to the cancellations that occur in the torus partition function as $n \to 1$. For $n\neq1$, the operators have different dimensions, meaning that there is no cancellation anymore and the expansion \eqref{eq:global_exp} contains terms of order $\mathcal{O}(1)$ at $n=1+\mathcal{O}(\delta)$ that are invisible if we only know the CFT data of the unitary subsector at $n = 1$. The fact that as we move from $n=1$ to $n=1+\delta$ new operators appear with $\ord(1)$ coefficients makes it unlikely that one can determine observables in the $n=1+\mathcal{O}(\delta)$ theory starting from just the Ising subsector of the $n=1$ theory.

A concrete example is provided by the global primaries $\veps'$ with dimension $(h_{\veps'},h_{\veps'})$ and $\psi$ with dimension $(h_{\veps}+2,h_{\veps}+2)$. In the limit $\delta \to 0$ they both have dimension $(\tfrac52,\tfrac52)$, and their OPE coefficients become equal and opposite,\footnote{We use $\lambda_{ijk}$ for OPE coefficients of Virasoro primaries, while $C_{ijk}$ for OPE coefficients of global primaries.}
\begin{equation}
\begin{aligned}
	C_{\veps \veps \veps'}^2&\simeq -0.2302 +\mathcal{O}(\delta)\,,\\
	C_{\veps \veps \psi}^2&\simeq 0.2302 +\mathcal{O}(\delta)\,.
\end{aligned}
\end{equation}
Therefore in the $n \to 1$ limit we obtain
\begin{equation}
	\lim_{\delta\to0}\Bigl(C_{\veps \veps \veps'} F_{\Delta_{\veps'},\ell_{\veps'}} + C_{\veps \veps \psi} F_{\Delta_{\psi},\ell_{\psi}}\Bigr) = 0\,,
\end{equation}
despite the fact that the OPE coefficients $C_{\veps \veps \veps'}$ and $C_{\veps \veps \psi}$ are both of order $\mathcal{O}(1)$ at $\delta =0$. This is an example of operators that do not appear in this correlation function at $\delta =0$, but that appear with coefficients of order $\mathcal{O}(1)$ as soon as $\delta \neq 0$. This shows the difficulty of continuing this correlation function from $n=1$ to $n=1+\ord(\delta)$ using only the data of the $n=1$ theory.

Finally, we would like to mention that the $O(n)$ model is not the only theory that exhibits these features. We could equally well have discussed the two-dimensional Potts model, which for our purposes displays the same qualitative physics. The $Q$-state Potts model in $d=2$ can be defined for $Q\in \mathbb{R}_{>0}$ and is critical for $Q\le4$. The theory is non unitary for generic $Q$, and in the $Q\to 2$ limit we obtain the Ising model as a subsector of the theory. Although the $Q\to 2$ and $n\to1$ theories share the same Ising subsector, it can be checked explicitly that they are distinct theories with different operator content. We would also like to stress that, by the same reasoning, while the two-dimensional $O(n\to 1)$ model (or the two-dimensional $Q\to 2$ Potts model) and the $d\to 2$ WF CFT share the same unitary subsector, there is no reason to expect that these theories coincide beyond this sector.

\section{The $d\to 2$ limit of WF} \label{sec:WF_2d}

Let's go back to the WF fixed point to focus on the original question we asked. We have seen that, because of multiplet recombination, an operator $W^{\mu \nu \rho}$ must exist for $d>2$ and be the descendant of the lightest $\mathbb{Z}_2$ even spin-4 operator $T^{\mu \nu \rho \sigma}$. As such, it must have dimension $\Delta \simeq 5$ for $d\gtrsim2$. What is the fate of this operator $W^{\mu \nu \rho}$ as $d\to2$?

Given the previous discussion about the lack of an operator with $\Delta = 5$, $\ell =3$ in the two-dimensional Ising model, we can rule out the idea that the $d \to 2$ limit of WF coincides with the two-dimensional Ising model. One possible scenario is the same one that we saw in the $O(n)$ model, with the theory recovering a unitary Ising subsector; while the spin-4 operator $T^{\mu \nu \rho \sigma}$ belongs to this subsector, the operator $W^{\mu \nu \rho}$ does not. We will now look for evidence of this scenario.\footnote{We do not consider the less generic scenario in which the theory splits into two decoupled subsectors, as this requires further conditions. Besides, for the two-dimensional $O(n)$ model, this scenario is not realized.}

\subsection{Negative multiplicity operators}

In the two-dimensional $O(n)$ model, we have seen that, in the $n\to1$ limit, only a small part of the theory becomes the Ising subsector. Operators that belong to $O(n)$ representations that don't make any sense for $n=1$, e.g. representations with negative multiplicity at $n=1$, cannot be in the Ising subsector and must drop out from correlation functions and from the partition function. This generically happens because these operators cancel against other operators with positive multiplicity.

In the scenario we propose for the $d \to 2$ limit of the WF CFT, the same must happen. As we change the number of dimensions, the symmetry group changes, and operators transforming in representations that make no sense for $d=2$ should not be part of the Ising subsector of the theory.
Given that their multiplicity is negative, they should cancel against other operators with positive multiplicity.

This could be the answer to the question we outlined at the beginning of the paper: what is the fate of the operator $W^{\mu\nu\rho}$, which is a descendant of the lightest spin-4 operator for $d>2$ but is not part of the $2d$ Ising model? A possible solution is that the operator $W^{\mu\nu\rho}$ drops out of Ising observables by canceling out against an operator with negative multiplicity. Given that the multiplicity of a spin-3 operator at $d=2$ is 2, 
a good candidate should be an operator that has multiplicity $-2$ for $d=2$.

Operators of a CFT transform in representations of the spacetime symmetry, i.e. the conformal group. In Euclidean space, this is $SO(d+1,1)$. Operators are labeled by their scaling dimension and by how they behave under rotations, i.e. how they transform under $SO(d)$.
If we were interested in irreps of $SO(d)$ for generic $d$, we'd run into a known problem: we don't know how to continue $SO(d)$ to a non-integer number of dimensions \cite{Binder:2019zqc}. While we know how to continue the invariant tensor $\delta_{ij}$ to non-integer $d$, we do not know how to do the same for the Levi-Civita tensor, because we don't know how to make sense of an object with a non-integer number of indices. Indeed, continuing irreps of $SO(d)$ across dimensions proves tricky, because for some special values of integer $d$ special things might happen. An example is the $SO(d)$ two-index antisymmetric representation of $A_{[i j]}$, labeled by the Young tableau $(1,1)$, which for generic $d$ has dimension $d(d-1)/2$. For $d=4$, however, it decomposes into two irreps of $SO(4)$, both of dimension 3. These are the self-dual and anti self-dual representations
\begin{equation}
	A^\pm_{ij}\veps_{ijkl} = \pm A^\pm_{kl}\,.
\end{equation}
In the language of $so(4) \simeq su(2)\oplus su(2)$, this decomposes into $(1,0)\oplus (0,1)$. This illustrates a general issue in studying a generic QFT in non-integer dimensions. An analytic continuation in $d$ is bound to miss such special cases.

However, we are in luck because the WF fixed point is parity invariant. This symmetry extends the symmetry group to $O(d) \cong SO(d) \rtimes \mathbb{Z}_2$. 
Then we do not need to worry about the Levi-Civita tensor, and $O(d)$ can be continued to non-integer $d$ \cite{deligne,Binder:2019zqc}.

We label $O(d)$ representations by their Young tableaux $(l_1,l_2,\ldots,l_n)$, with $n$ rows and the $k$-th row consisting of $l_k$ boxes, with $l_{i}\ge l_{i+1}$. For example, a spin $\ell$ field is transforming in the $(\ell)$ irrep. Given an $O(d)$ irrep labeled by a Young tableau, there is an algorithm for computing its dimension which can be found in \cite{King_1971}; see also \cite[Section 4.3]{Bekaert:2006py}.

We will be particularly interested in operators that transform in irreps that have negative dimensionality at $d=2$. An example is the representation $(2,2)$, which has multiplicity $M_{2,2}$
\begin{equation}
\yng(2,2)\,, \qquad \qquad  M_{2,2} = \frac{1}{12} (d-3) d (d+1) (d+2) 
\end{equation}
This is well defined for integer $d>3$, becomes null at $d=3$ and has multiplicity $-2$ at $d=2$.
The same happens for $(k,2)$,
\begin{equation}
	\overbrace{\yng(5,2)}^\text{$k$ boxes}\,, \qquad  M_{k,2} = (d-3) d (k-1) (d+k-1) (d+2 k-2) \frac{\Gamma (d+k-3)}{2 \Gamma (d-1) \Gamma (k+2)}\,.
\end{equation}
For all allowed values of $k$, we have $M_{k,2}|_{d=2}=-2$, $M_{k,2}|_{d=3}=0$ and $M_{k,2}|_{d=4}>0$. As discussed for the $O(n)$ model, when we analytically continue a theory to $d \in \mathbb{R}$, we cannot simply discard these operators, despite the fact that their multiplicity makes no sense for integer $d \le 3$.

Negative multiplicity by itself is not enough to explain the cancellation of $W^{\mu \nu \rho}$.
A further requirement of this scenario is that such a negative-multiplicity operator has the same dimension as $W^{\mu\nu\rho}$ at $d=2$, meaning $\Delta = 5$. This is a condition that can be checked explicitly in the epsilon expansion, given that anomalous dimensions can be computed to high order.
Here we just show that the one-loop computation in the $4-\eps$ expansion gives a result in the right ballpark for some of these operators, and we postpone a more systematic and higher-order computation in the $\eps$ expansion to future work.

\subsection{Scaling dimensions}
We are interested in the scaling dimensions of operators transforming in the $(k,2)$ representation of $O(d)$. They should also be $\mathbb{Z}_2$ even, given that $\mathbb{Z}_2$ is a symmetry for any value of $d$ and they need to appear in the same correlation functions in order to cancel. Finally, we will only focus on the lightest operator transforming in a given representation of $O(d)$.

Let's first consider an operator transforming in the $(2,2)$ representation. One-loop scaling dimensions of such operators can be found in \cite{Kehrein:1994ff}; see Table 4 and Equations (14)-(15). At one loop, the scaling dimension of an operator transforming in this representation with $n$ powers of the field $\phi$ is\footnote{The work \cite{Kehrein:1994ff} identifies operators by their $so(4) \cong su(2)\times su(2) $ representations. The operator we are interested in is the $(2,0)+(0,2)$ one in Table 4 of that work.}
\begin{equation}
\Delta_{\rm WF} = 4 +n\left(\frac{d}{2}-1 \right) + \frac{\epsilon}{18} (n-3)(3n+2)+O(\epsilon^2)
\end{equation}
which means that, for $d=\epsilon=2$ we have
\begin{equation}
\left. \Delta_{\rm WF} \right|_{d=2} = 4 +\frac{(n-3)(3n+2)}9+\ldots
\end{equation}
The lightest operator that is $\mathbb{Z}_2$ even has $n=4$, and we find
\begin{equation}
	\left. \Delta_{\rm WF} \right|_{d=2} = 4 +\frac{14}9+\ldots \simeq 5.56
\end{equation}
Higher values of $n$ extrapolate to larger values of $\Delta$ at $d=2$.  

We can also look at other operators transforming in the $(k,2)$ representation. For a few values, their one-loop anomalous dimensions are collected in Table \ref{tab:1loop}.
\begin{table}
\begin{center}\bgroup
\def\arraystretch{2.5}
\begin{tabular}{ L | C | C | L }
O(d) \text{ irrep} & so(4)\cong su(2)\times su(2) \text{ irrep } & \Delta_{\rm free}|_{d=2} & \text{One-loop } \Delta_{\rm WF}|_{d=2} \\
\hline
\yng(2,2) & (2,0)+(0,2)& 4& 4 +\frac{14}9 \simeq 5.56^\text{\cite{Kehrein:1994ff}} \\ 
\yng(3,2) & (5/2,1/2)+(1/2,5/2)&  5&  5+\frac49\simeq 5.44^\text{\cite{Kehrein:1994ff}} \\ 
\yng(4,2) & (3,1)+(1,3)& 6& 6+\frac{1}{90} \left(37 \pm \sqrt{649}\right) \simeq 6.26, 7.39^\text{\cite{Henriksson:2022rnm}} \\ 
\yng(5,2) &(7/2,3/2)+(3/2,7/2)& 7 & 7+\frac{1}{60} \left(27 \pm \sqrt{129}  \right) \simeq 7.52, 8.28 \\ 
\end{tabular}
\egroup
\end{center}
\caption{Results for the one-loop scaling dimensions of the lightest $\mathbb{Z}2$-even operator in a given $O(d)$ representation. In the UV, the operators have scaling dimension $\Delta{\rm free}$, while in the IR they have scaling dimension $\Delta{\rm WF}$. All operators are built from $n=4$ powers of the field $\phi$ and $\Delta_{\rm free}|_{d=2}$ derivatives.}
\label{tab:1loop}
\end{table}

From these results, it appears that the lightest operators transforming in the $(2,2)$ and in the $(3,2)$ representations have scaling dimensions in the right ballpark to cancel out with the dimension 5 operator $W^{\mu \nu \rho}$. 
Such a cancellation should be observed in correlation functions where both operators appear, such as a four-point function with four spin-2 operators for an operator transforming in the $(2,2)$ representation, or a four-point function with two spin-2 operators and spin-3 operators for the operator transforming in the $(3,2)$ representation.
Computing higher order corrections to the lowest dimensional operator transforming in these representations will tell us which is the most likely candidate to cancel against $W$.\footnote{In general, every operator transforming in an irrep with negative multiplicity must be canceled by operators with positive multiplicity. This requirement applies to all operators listed in Table \ref{tab:1loop}. As seen in Section \ref{ssub:correlation_functions_of_non_ising_operators}, we cannot identify on first principles what operators they should cancel against. Further computations e.g. in the epsilon expansion are therefore needed in order to find the most promising candidates to cancel operators such as $(4,2)$ and $(5,2)$ in Table \ref{tab:1loop}.}

\subsubsection{$d=3$}
The mechanism we have illustrated can hold in general for any limit in which $d$ approaches an integer. We have focused on $d \to 2$, because the knowledge of the full $d=2$ Ising spectrum makes some paradoxes manifest. However, the same discussion about operators transforming in representations with negative multiplicity can be applied to the $d\to 3$ limit of the Wilson-Fisher fixed point. Concretely, the scenario predicts that for every operator with negative multiplicity in the $d = 3$ limit, there should be at least another operator with positive multiplicity that has the same scaling dimension. This can be checked in the epsilon expansion.

$O(d)$ representations that have negative multiplicity at $d=3$ are represented by Young tableaux with at least three rows; the simplest examples are $(k,2,1)$ and $(k,2,2)$. Operators transforming in these representations are therefore expected not to be part of the Ising subsector, and to drop out of correlation functions by canceling against other operators. Given that the multiplicities of $O(d)$ representations can have zeros only at integer values of $d$, for a representation to have negative multiplicity at $d=3$ it means that it must have either negative or vanishing multiplicity at $d=4$. These are therefore operators that might be easily missed when studying the theory in $d=4-\eps$ dimensions, but are still part of the spectrum of the WF CFT.

\subsection{From $d=2$ to $d=2.01$}  \label{sec:d=2.01}
Another question we are ultimately interested in is whether knowledge of the conformal data of the WF CFT in integer $d$ spacetime dimensions allows us to continue the conformal data to $d+O(\epsilon)$ dimensions. After all, one might hope that this is feasible, given the assumption that the conformal data depends analytically on the spacetime dimension. More specifically, we would like to know if knowledge of the Ising model at integer $d$ is enough to compute scaling dimensions for the WF CFT in $d+\ord(\eps)$ dimensions.

Around $d=4$ the WF fixed point is weakly coupled, so the perturbative description \eqref{eq:WF_action} allows one to compute conformal data in $d=4-\epsilon$ dimensions. But without explicitly using the weakly coupled description, would it be possible to obtain the same results? It was shown that scaling dimensions can be computed at leading order in $\epsilon$ in \cite{Rychkov:2015naa}, and up to order $\epsilon^3$ in \cite{Gopakumar:2016wkt,Gopakumar:2016cpb}, using purely CFT techniques. However, going beyond this order has not been achieved.

What about $d=2$ or $d=3$? Can the exact two-dimensional results or numerical three-dimensional results for the conformal data of the Ising model be used to study the WF fixed point in $d=2 +\epsilon$ or $d=3 \pm \epsilon$ dimensions? The discussion of the previous section suggests that this might not be possible. In the scenario we discussed, the Ising model in $d=2$ dimensions is only a subsector of the $d\to 2$ limit of the WF fixed point, so it is unlikely that we can describe the full $d=2+\epsilon$ theory starting from knowledge of just a subsector of the theory in the $d \to 2$ limit. Indeed, while a $d=2+\eps$ expansion has been worked out for some models, the same has not been achieved with the Ising model.

A famous example of a continuation to $d=2+\eps$ dimensions is that of the $O(n>2)$ nonlinear sigma model \cite{Polyakov:1975rr}. The situation here is different from the case at hand, since we have a lagrangian description of the theory. From this point of view, this appears to be a situation more similar to the WF fixed point in $d=4-\eps$ dimensions, and results up to three loops have been computed \cite{Hikami:1977vr}.\footnote{While it is often believed that the interacting fixed point is in the same universality class as the $O(n)$ WF model, this has recently been questioned by pointing out that the $2+\eps$ expansion has conserved operators that are not expected in the $O(n)$ WF CFT \cite{Jones:2024ept,DeCesare:2025ukl}. Proposed solutions to this problem seem to exclude the scenario where we have both analytic behavior in $d$ for $2<d<4$ for the WF $O(n)$ universality class and agreement between the NLSM and the WF $O(n)$ model at $d=2+\eps$ \cite{DeCesare:2025ukl}.} Other expansions to $d=2+\eps$ dimensions have been computed, such as the Gross-Neveu model \cite{Kivel:1993wq} and the random-bond Ising model \cite{Komargodski:2016auf}; both of these cases have a perturbative formulation. 

For what concerns the Ising model, the author of \cite{Li:2021uki} attempted to find numerical solutions to the crossing equations at leading order in a $d-2$ expansion, without success. The main assumptions behind the computation were that around $d=2$ all conformal data behaves linearly in $d-2$, and that in any correlation function of $\sigma$ and $\veps$ operators, all new states have OPE coefficients of order $\ord(d-2)$, and therefore appear in crossing equations at subleading order. The fact that this assumption does not allow a solution for $d\gtrsim2$ agrees with our scenario because, as seen explicitly in Section \ref{sec:O(n=1.01)}, we expect new operators to appear with $\ord(1)$ coefficients, thanks to the non-unitarity of the theory in non-integer dimensions. %

A bootstrap approach to the Ising universality class in non-integer dimensions has been carried out in \cite{El-Showk:2013nia,Cappelli:2018vir,Henriksson:2022gpa}. These studies do not rely on perturbative expansions around $d=4$ or $d=2$ but, on the other hand, are necessarily non-rigorous, given that the bootstrap approach assumes unitarity. Despite this, the method appears to give reasonable results. We point out that these studies have used correlation functions of the scalars $\sigma$ and $\veps$, which only exchange irreps of $O(d)$ described by Young tableaux with a single row. These irreps of $O(d)$ are well behaved in the limits $d\to2,3$. To observe operators that have negative multiplicity as $d\to2$, one should study correlation functions of operators with spin. 
However, as we have learned from vector operators in the $O(n)$ model, it should still be possible to see some operators decoupling from the Ising sector of the theory. This phenomenon was suggested to take place in \cite{Henriksson:2022gpa}, where a numerical study in $d=2.001$ points to evidence for a scalar operator decoupling in some way from the theory. Finally, a multicorrelator analysis of the Ising model in $2<d<3$ shows the island of allowed parameter space shrinking to zero as we approach $d=2$ \cite{Ning_unp}.

Finally, we mention another example, similar in spirit to the present work, that has been discussed in the literature: the extension of WZW models to non-integer dimensions. In \cite{Zhou:2022pah}, the phenomenon of recombination of higher spin conserved currents for $d\gtrsim 2$ in a fermionic theory that in the IR is described by the $SU(N)_1$ WZW model (e.g. the $N$-flavor Schwinger model \cite{Gepner:1984au,Affleck:1985wa,Dempsey:2023gib,Mouland:2025ilu}) was discussed. As a solution, it was proposed that, as $d \to 2$, new currents emerge because the two-point function of the field strength goes to zero. We'd like to point out that this is not a sufficient condition for the decoupling of an operator: the overall normalization of the field can always be redefined so that it's $\ord(1)$ in the $d\to2 $ limit. One would need to look at other correlation functions of the field strength with other operators as well. As an example, in a free bosonic theory with the standard normalization $\braket{\phi(x)\phi(0)} = |x|^{-2\Delta}$, the two point function of $\partial_\mu \phi$ vanishes in the $d\to2$ limit; however, this operator is still a good operator at $d= 2$.

\section{Conclusions}
In this work, we discussed the identification between the Wilson-Fisher CFT in the limit of integer dimensions and the Ising CFT. Agreement of the lightest scaling dimensions in the two theories might suggest that these two theories should be identified with each other. We have shown that this conclusion leads to a paradox, which is particularly sharp in the $d \to 2$ case. In this limit, the infinitely many conserved currents appearing due to Virasoro symmetry lead to the existence of infinitely many global primaries in the $d\to2$ limit of the Wilson-Fisher CFT that are not part of the Ising CFT.

We have therefore put forward a different scenario, in which the Wilson-Fisher CFT develops a unitary subsector which coincides with the Ising CFT. Correlation functions inside this subsector reproduce exactly those of the Ising model. However, the full theory is larger than the Ising CFT, and contains many additional operators. Operators outside the Ising subsector must cancel out from correlation functions in which the external operators belong to the Ising subsector.

We have observed this phenomenon explicitly in the two-dimensional $O(n)$ model, which develops an Ising subsector as $n\to1$. In this case, thanks to the knowledge of the scaling dimensions and some OPE coefficients, it's possible to explicitly observe the cancellation between non-Ising operators in Ising correlation functions. These cancellations happen between operators with positive multiplicity and operators with negative multiplicity, i.e. operators transforming in irreps of $O(n)$ that have negative multiplicity for $n=1$ but still need to be considered for $n \in \mathbb{R}$ and, therefore, in the $n \to 1$ limit.
Using this intuition, we have conjectured that the Wilson-Fisher operators that do not become Ising operators in the $d\to2$ limit should also cancel against negative multiplicity operators. %
This scenario would mean that the scaling dimension of these operators must match; this can be checked for example in the epsilon expansion. We have identified the most likely candidates, the lightest primary transforming either in the $(2,2)$ or in the $(3,2)$ representation of $O(d)$, and shown that their one-loop scaling dimensions are in the right ballpark to cancel out some of the unwanted operators.

Finally, we have discussed how this scenario makes it unlikely that one can continue the Ising conformal data from integer dimensions to non-integer dimensions, a task which indeed has not been achieved so far. It looks like in order to describe the theory in e.g. $d=2+\eps$ dimensions we would need knowledge of the full $d\to 2$ limit of the Wilson-Fisher fixed point, rather than just its Ising subsector.

Studying the full $d\to2$ theory appears to be a tricky task. The theory is most likely logarithmic: cancellation between different operators means that one can take an appropriately rescaled linear combination of the two and obtain correlation functions which are finite and logarithmic. This is the usual mechanism of \cite{Cardy:2013rqg}. Furthermore, the lack of unitarity means that the usual BPZ approach \cite{Belavin:1984vu} fails: operators in the unitary sector have zero-norm descendants whose correlation function with operators outside of the unitary sector does not need to vanish. These zero-norm descendants therefore cannot be used to study mixed correlation between operators inside and outside the unitary subsector. However, the full extended theory should still satisfy the usual constraints of non-unitary CFTs, and be for example crossing symmetric.

Several directions remain open. In order to concretely test the scenario further, the computation of scaling dimensions of the negative multiplicity operators should be carried out beyond one loop; the techniques of \cite{Henriksson:2025vyi} seem well-suited to this task. Given the precision that can be achieved through the epsilon expansion, this could give extremely strong evidence for or against this scenario.
The same question could also be analyzed in other theories that are expected to exist in non-integer dimensions, such as the $O(n)$ Wilson-Fisher fixed point. %
More generally, we currently don't have a method to obtain the conformal data of a CFT at $d+\eps$ starting from the known conformal data at $d$, with $d \in \mathbb{R}$, that does not rely on a perturbative definition; if such a method could be developed, it would prove to be a handle on the WF fixed point.
Finally, we have seen that the same unitary subsector can be embedded in different non-unitary CFTs, such as the two-dimensional $O(n\to1)$ and $S_{Q\to2}$
 models. An interesting question is whether the full non-unitary theory can be fixed directly in two dimensions and, if so, which additional physical input (for example, the presence of a spin $3$, dimension $5$ operator) are required to do so. If this were possible, it would remove one of the main obstacles to developing a $d=2+\eps$ expansion.

\section*{Acknowledgements}
The author thanks Leonardo Rastelli for collaboration at the early stages of this project, for several discussions and for comments on the draft. The author thanks Connor Behan, Johan Henriksson and Ning Su for discussions, and Slava Rychkov for extended discussions and comments on the draft.

\appendix
\section{Details about the $O(n)$ model} \label{app:on}
\subsection{Partition function}
The partition function of the $O(n)$ model for generic $n$ is given in \eqref{eq:ZOn}. It depends explicitly on \cite{Read:2001pz}
\begin{equation} \label{eq:Lambda}
	\Lambda_{M, N} = 
2 \sum_{d > 0 : d | M} 
\frac{\mu\!\left(\frac{N}{N \wedge d}\right) \phi(M/d)}
{M \, \phi\!\left(\frac{N}{N \wedge d}\right)}
\cos(\pi e_0 d)
\end{equation}
with $\phi(m)$ the Euler totient function
\begin{equation}
	\phi(m)=m \prod_{\substack{p\text{ prime}\\p|m}}(1-p^{-1})
\end{equation}
and $\mu(m)$ is the Möbius function, which has the value $\mu(m) = (-1)^k$ if $m$ is the product of $k$ distinct prime numbers, $\mu(1) = 1$, and $\mu(m) = 0$ otherwise.

\subsection{Correlation functions}
Computing correlation functions of the $O(n)$ model is a relatively painless task if we look at correlation functions that include the energy operator $\veps$. Given that it's a degenerate operator with dimensions $(h_{1,3},h_{1,3})$, and its level three descendant is null, these correlation functions satisfy a third order ODE.

As in the main text, we parametrize the central charge of the theory as
\begin{equation}
	c = 1-\frac{6}{m(m+1)}
\end{equation}
and the null descendant is
\begin{equation}
	\left(L_{-3}-\frac{2 (m+1)}{3 m+1}L_{-1}L_{-2}+\frac{(m+1)^2}{2 m (3 m+1)}L_{-1}^3\right)\veps = 0
\end{equation}
meaning that correlation functions satisfy the BPZ differential equation \cite{Belavin:1984vu}
\begin{equation}
	\left(\calL_{-3}-\frac{2 (m+1)}{3 m+1}\calL_{-1}\calL_{-2}+\frac{(m+1)^2}{2 m (3 m+1)}\calL_{-1}^3
	\right)
	\braket{\veps(z)\calO_1(z_1)\ldots\calO_N(z_N)}=0
\end{equation}
with the differential operators $\calL_j$ defined by
\begin{equation}
	\mathcal{L}_{-k} = \sum_{i=1}^{N} 
	\left[ \frac{(k - 1)h_i}{(z_i - z)^k} - \frac{1}{(z_i - z)^{k-1}} \frac{\partial}{\partial z_i} 
\right]\,.
\end{equation}

In the case of a four-point function, solutions to the BPZ differential equations are the Virasoro blocks $\calV_i$
\begin{equation}
 	\braket{\veps(0) \veps(z,\bz) \calO(1) \calO(\infty)} = \sum_i \lambda_{\veps \veps i}\lambda_{\calO \calO i} \calV_{i}(z)\overline{\calV}_{i}(\bz)= \sum_i \lambda_{\veps \calO i}^2 \tilde{\calV}_{i}(1-z)\overline{\tilde{\calV}}_{i}(1-\bz)\,,
 \end{equation} 
 where we have included both the $s$-channel and the $t$-channel expansion.
The fact that the BPZ equation is of order three means that the operators exchanged can only have one of three holomorphic weights. Solving the BPZ equation to leading order shows that the weights allowed in the fusion of an operator with weight $h_{1,3}$ and one with $h_{r,s}$ are
\begin{equation}\label{eq:13_fusion}
 	h_{1,3} \times h_{r,s} = h_{r,s-2}+h_{r,s}+h_{r,s+2}\,.
 \end{equation}
This is the well known fusion rules for minimal models, but holds for non-integer $r$ and $s$ as well.

It's generically impossible to solve this differential equation in closed form. Our strategy is therefore to compute the Virasoro blocks for some value of $m$ to a very high order in an expansion in $z, \bz$ for the s-channel and in $1-z,1-\bz$ for the t-channel. This can be easily done to $z^{\calO(100)}$ in Mathematica. Then we obtain the 
 OPE coefficients by imposing crossing symmetry at and around $z=\bz=1/2$.
 Here we report some results for the correlation functions we mention in the main text.

 \subsubsection*{$\braket{\veps\veps\veps\veps}$}
 This correlation function only contains degenerate scalar operators, and by the fusion rules \eqref{eq:13_fusion} it can only exchange operators with zero-spin. This fixes this correlation function in the $O(n)$ model to be exactly the same as in the generalized minimal model with the same central charge. We do not need to do any further computation, and we can just read off the OPE coefficients from \cite{Dotsenko:1984ad} to be (the function $C^{\rm MM}$ are those defined in \cite[Equation (A.5)]{Poghossian:2013fda})
 \begin{equation} \label{eq:eps_OPE}
 \begin{aligned}
 	\lambda_{\veps \veps \veps}&=C^{\rm MM}_{(1,3),(1,3),(1,3)}\,, \\ 
 	\lambda_{\veps \veps \veps'}&=C^{\rm MM}_{(1,3),(1,3),(1,5)}\,.
 \end{aligned}
 \end{equation}

\subsubsection*{$\braket{\veps \veps V V}$}
The operator $V$, which transforms in the vector representation of $O(n)$, has weights $(h_{1/2,-2},h_{1/2,2})$ and spin $h_{1/2,-2}-h_{1/2,2}=1$; it is not a degenerate operator, therefore we cannot read the OPE coefficients from those of generalized minimal models.

However, once we fix the OPE coefficients using the procedure described above, for several values of $m$ we observe that the OPE coefficients are closely related to those of minimal models. We report here only the OPE coefficients appearing in the s-channel
\begin{equation}
\begin{aligned}
	\lambda_{VV\veps} &= 
	\left( C^{\rm MM}_{(1,3),(1/2,2),(1/2,2)}C^{\rm MM}_{(1,3),(1/2,-2),(1/2,-2)}\right)^{1/2}\,,
	\\
	\lambda_{VV\veps'} &= -
	\left( C^{\rm MM}_{(1,5),(1/2,2),(1/2,2)}C^{\rm MM}_{(1,5),(1/2,-2),(1/2,-2)}\right)^{1/2}\,.
\end{aligned}
\end{equation}
Once we have obtained the Virasoro blocks and the OPE coefficients for generic values of $m$, we can take the $m \to 3$ limit.
It can be observed that, in this limit, the product $\lambda_{\veps \veps \veps} \lambda_{VV\veps}$ vanishes while $\lambda_{\veps \veps \veps'} \lambda_{VV\veps'}$ blows up, but the correlation function remains finite. We recover a correlation function which exchanges logarithmic operators, arising from the fact that, for $m \to 3$, we have $h_{\veps'}-h_{\veps} =2$. Given that the dimension of $\veps'$ concides with the dimension of a level two descendant of $\veps$, the two can and do form a logarithmic multiplet, which in this case is staggered \cite{Gaberdiel:1996kx,Rohsiepe:1996qj}.

\subsubsection*{$\braket{\veps \veps Y Y}$}
The situation is the same for the operator $Y$, which has dimensions $(h_{3/2,-2/3},h_{3/2,2/3})$, and spin 1. OPE coefficients are found to be again closely related to the minimal model ones
\begin{equation}
\begin{aligned}
	\lambda_{YY\veps} &= -\left( C^{\rm MM}_{(1,5),(3/2,2/3),(3/2,2/3)}C^{\rm MM}_{(1,5),(3/2,-2/3),(3/2,-2/3)}\right)^{1/2}\,,
	\\
	\lambda_{YY\veps'} &= \left( C^{\rm MM}_{(1,5),(3/2,2/3),(3/2,2/3)}C^{\rm MM}_{(1,5),(3/2,-2/3),(3/2,-2/3)}\right)^{1/2}\,.
\end{aligned}
\end{equation}
In the $m \to 3$ limit, the situation is similar to that of the $\braket{\veps \veps VV}$ correlation function, with one OPE coefficient going to zero and the other one diverging. The $m \to 3$ limit of $\braket{\veps \veps YY}$ still remains finite, and one can check explicitly that it agrees with $\braket{\veps \veps VV}$
\begin{equation}
	\lim_{m \to 3} \braket{\veps \veps YY}-\braket{\veps \veps VV} = 0\,.
\end{equation}
This means that we can build a logarithmic multiplet out of $V$ and $Y$ in this limit. Reinstating indices for a moment, we can build the following combination which has finite correlation functions with $\veps$ in the $n \to 1$ limit,
\begin{equation}
Y_{abc} +\frac{1}{\sqrt{2(n-1)}}\left( 2\delta_{ab} V_c-\delta_{ac} V_b-\delta_{bc} V_a \right)\,,
\end{equation}
with $Y_{abc}$ symmetric under $ a \leftrightarrow b$. Correlation functions of this operator will be logarithmic.

\subsubsection*{$\braket{\veps \veps J J}$}
The current operator $J$, of weights $(1,0) = (h_{1,-1},h_{1,1})$, is not a degenerate operator in its holomorphic part, and therefore like earlier we cannot use directly results from the minimal models. This correlation function was studied explicitly in \cite{Gorbenko:2020xya}, and it has the further peculiarity that logarithmic operators are exchanged in the $\veps \times J$ OPE at generic values of $n$. This does not change the method in which we fix the OPE coefficients, and we find
\begin{equation}
\begin{aligned}
	\lambda_{JJ\veps} &= -\lim_{\tau \to 0}\left( C^{\rm MM}_{(1,3),(1,1+\tau),(1,1+\tau)}C^{\rm MM}_{(1,3),(1,-1-\tau),(1,-1-\tau)}\right)^{1/2}\,,\\
	\lambda_{JJ\veps'} &= -\lim_{\tau \to 0}\left( C^{\rm MM}_{(1,5),(1,1+\tau),(1,1+\tau)}C^{\rm MM}_{(1,5),(1,-1-\tau),(1,-1-\tau)}\right)^{1/2}\,.
\end{aligned}
\end{equation}

\bibliographystyle{utphys}
\bibliography{biblio}

\end{document}